\documentclass[12pt]{article}
\usepackage{latexsym,graphicx,multirow}
\usepackage{float}
\usepackage{amssymb}
\usepackage{amsmath}
\usepackage{amscd}
\usepackage{amsthm}
\usepackage[left=2cm,top=2.5cm,right=2.5cm,bottom=1.5cm]{geometry}
\usepackage{hyperref}
\usepackage{epstopdf}
\usepackage{cite}

\begin{document}
	
	\begin{center}
		\large{\bf{ Barrow HDE model for Statefinder diagnostic in FLRW Universe }} \\
		\vspace{10mm}
		\normalsize{ Anirudh Pradhan$^1$, Archana Dixit$^2$, Vinod Kumar Bhardwaj$^3$  }\\
		\vspace{5mm}
		\normalsize{$^{1,2,3}$Department of Mathematics, Institute of Applied Sciences and Humanities, GLA University\\
			Mathura-281 406, Uttar Pradesh, India}\\
		\vspace{2mm}
		$^1$E-mail: pradhan.anirudh@gmail.com \\
		\vspace{2mm}
         $^2$E-mail: archana.dixit@gla.ac.in\\
         \vspace{2mm}
         $^3$E-mail: dr.vinodbhardwaj@gmail.com\\

\vspace{10mm}
	
%\date{}
%\maketitle

 \end{center}
\begin{abstract}
	
We have analyzed the Barrow holographic dark energy (BHDE) in the framework of the flat FLRW Universe by considering the various estimations 
 of Barrow exponent $\triangle$. Here we define BHDE, by applying the usual holographic principle at a cosmological system, for utilizing 
 the Barrow entropy rather than the standard Bekenstein-Hawking. To understand the recent accelerated expansion of the universe, considering 
 the Hubble horizon as the IR cut-off. The cosmological parameters, especially the density parameter ($\Omega_{_D}$), the equation of the state 
 parameter ($\omega_{_D}$), energy density ($\rho_{_{D}}$) and the deceleration parameter($q$) are studied in this manuscript and found 
 the satisfactory behaviors. Moreover, we additionally focus on the two geometric diagnostics, the statefinder $(r,s)$ and $O_{m}(z)$ to discriminant 
 BHDE model from the $\Lambda CDM$ model. Here we determined  and plotted the trajectories of evolution for statefinder $(r, s)$, $(r,q)$ and 
 $O_{m}(z)$ diagnostic plane to understand the geometrical behavior of the BHDE model by utilizing Planck 2018 observational information.
Finally, we have explored the new Barrow exponent $\triangle$, which strongly affect the  dark energy equation of state that can lead it 
to lie in the quintessence regime, phantom regime, and exhibits the phantom-divide line during the cosmological evolution.
\end{abstract}
 
 
 %%%%%%%%%%%%%%%%%%%%%%%%%%%%%%%%%%%%%%%%%%%%%%%%%%%%%%%%%%% Section 1 %%%%%%%%%%%%%%%%%%%%%%%%%%%%%%%%%%%%%%%%%%%%%%%%%%%%%%%%%
 
\section{Introduction}

The well-proven accelerated expansion of the universe is the greatest achievement of $20^{th}$ century \cite{ref1,ref2}. The dark energy (DE) 
with immense negative pressure is considered one of the mysterious reasons behind the accelerated expansion of the universe. The WMAP 
experiment also suggests that the Universe is made up of $4\%$ of the Baryonic matter, $23\%$ DM and $73\%$ DE \cite{ref3}. In the path of 
expansion, the universe passes through different phases of DE/matter. The DE is typically defined by the EoS parameter $(\omega)$  and the 
ranges include $-1/3<\omega<-1$ for quintessence, $\omega<-1$ for phantom and $(\omega =1)$ for the cosmological constant.\\
  
Recently, researchers show a great interest in HDE models, since these HDE models developed as applications of DE by following 
holographic principle \cite{ref4}. The holographic principle derives from the thermodynamics of the black hole. String theory 
provides a relation between the IR cutoff of quantum field theory linked to vacuum energy  \cite{ref4}-\cite{ref7}. This concept 
has been utilized widely in cosmological contemplations, especially in the late-time period of the Universe, at present known as, 
holographic dark energy models \cite{ref8}-\cite{ref22}. During this phase, we would like to mention that  Nojiri-Odintsov cut-off \cite{ref8} 
gave the most general holographic dark energy and it is intriguing that it might be applied to covariant hypotheses \cite{ref23}. So for solving 
the dark energy puzzle, (HDE) speculation is a promising approach \cite{ref16}-\cite{ref18}. The new HDE models can be proposed by utilizing 
holographic speculation and a generalized entropy. In addition to the dark energy model, it is also found that the HDE is important to analyze 
the early evolution of the Universe, such as the inflationary evolution \cite{ref24}-\cite{ref29}. It is worth mentioned here some latest papers 
\cite{ref29a}-\cite{ref29f} and their references on HDE in various scenarios. \\

In this present work, we consider a spatially flat, homogeneous, and isotropic spacetime as the underlying geometry. Here we study the 
behavior of different cosmological parameters (the deceleration parameter, the energy density parameter, and the equation of state
parameter ) during the cosmic evolution by assuming the Hubble horizon as the infrared (IR) cut-off. The Hubble horizon as an IR cut-off 
is suitable to clarify the ongoing accelerated expansion of the DE models.\\

In this direction, many cosmologists have presented mathematical diagnostics ${r, s}$, known as statefinder parameters. For observing the 
nature of DE models, statefinder parameters are the most important parameter \cite{ref30}-\cite{ref31}. In order to discriminate the various 
DE models, the trajectories can be represented graphically in $r-s$ and $r- q$ planes. The state finder parameters are also 
analyzed \cite{ref32,ref33}. DE models like Chaplygin gas models, quintessence model, cosmological constant and braneworld model as 
explored in \cite{ref34} -\cite{ref38}.\\

Barrow holographic dark energy (BHDE) is also a fascinating alternative scenario for the quantitative description of DE which is based on 
the holographic hypothesis \cite{ref39}-\cite{ref43}  and applying the recently proposed Barrow entropy \cite{ref44} instead of the normal 
Bekenstein-Hawking \cite{ref45,ref46}. Saridakis \cite{ref47} have shown that the BHDE includes basic HDE as a sub-case in the limit where 
Barrow entropy becomes the usual Bekenstein-Hawking. Anagnostopoulos et al. \cite{ref48} have shown that the BHDE is an agreement with 
observational data, and it can serve as a good candidate for the description of DE. Barrow holographic dark energy models have been studied 
by several authors \cite{ref48a} -\cite{ref48f} in different contexts.

On the other hand, concerning various cosmological theories, 
where DE interacts with DM has extended much attention in the literature \cite{ref49}.\\ 

 The essential aspect in holographic principle in the cosmological level, is that the universe Horizon entropy is proportional to its area,
 as similar to the Bekenstein-Hawking entropy, with a black hole. The entropy of the black hole shown by Barrow  can be modified as \cite{ref44}
 \begin{equation}
 \label{1}
 S_{B}= \left(\frac{B}{B_{0}}\right)^{1+\frac{\triangle}{2}} ~~~~~0\le\triangle\le1 ,
 \end{equation}
 where  $B_{0}$ is the Planck area and $B$ is the normal horizon area.\\
 
 There is a quantum-gravitational deformation which enumerated by the parameter $\triangle$ , for $\triangle = 0$
 related to the Bekenstein-Hawking standard entropy and $\triangle = 1$ corresponding to the most complex and fractal structure.
 The aim of the present manuscript is to examine the BHDE model by taking the Hubble radius as an IR cutoff and analyzing the 
 behavior of cosmological parameters for a flat FRW universe. We extend our analysis to BHDE, inspired by the works \cite{ref47} with a similar 
 IR cut-off which gives the ongoing stage progress of the Universe. We get the statefinder parameters for BHDE which accomplish the 
 worth of $\Lambda CDM$  model and show consistency with the quintessence model for appropriate estimation of parameters. The plan of 
 this manuscript is as follow: In section $2$, we introduce the BHDE model proposed in \cite{ref47} with a general interaction term between 
 the dark components (BHDE and DM) of the universe and also study its cosmological evolution by considering the basic field equations. 
 The behavior of state finder pair for BHDE has been discussed in section $3$, we explore the $O_{m}$ diagnostic in section $4$. Finally, 
 section $5$ is devoted to conclusions.

%%%%%%%%%%%%%%%%%%%%%%%%%%%%%%%%%%%%%%%%%%%%%%%%%%%%%%%%%%%%%%%%%% Section 2 %%%%%%%%%%%%%%%%%%%%%%%%%%%%%%%%%%%%%%%%%%%%%%%%%%%%%%%%%%%%%%%%

\section {Basic field Equations} 

In this section we develop the scenario of Barrow holographic dark energy, where the inequality $\rho_{_{D}}L^{4}\leqslant S$, is given 
by the standard HDE. Here $L$ is the horizon length under the assumption $S \propto A \propto L^{2}$ \cite{ref9} by using the Barrow 
entropy (\ref{1}) obtain as lead to

\begin{equation}
\label{2}
\rho_{_{D}}= C L^{\triangle-2},
\end{equation}

where $C$ is a parameter with dimensions $[L]^{-2-\triangle}$ and $L$ denotes the IR cutoff. In the case where $\triangle = 0$ as expected, the above 
expression provides the standard holographic dark energy $\rho_{D} = 3c^{2} M_{p}^{2} L^{-2}$ (here $M_{p}$ is the Plank mass), where 
$C=3c^{2} M_{p}^{2}$ and with the model parameter $c^{2}$. The above relation leads to some interesting results in the 
holographic and cosmological setups \cite{ref47,ref48}. In \cite{ref49a,ref48b}, Barrow entropy was added in the structure
of “gravity-thermodynamics” conjecture, according to which the first law of thermodynamics can be applied on the universe apparent horizon. 
As a result, one obtains a modified cosmology, with extra terms in the Friedmann equations depending on the new exponent $\triangle$,
which disappear in the case $\triangle=0$, i.e when Barrow entropy becomes the standard Bekenstein-Hawking one. Although this framework is 
defined in a very effective way in the universe of late time. It should be noted here that the value $\triangle=1$ corresponds to the 
maximal deformation, while the value $\triangle=0$ corresponds to the simplest horizon structure, and the normal Bekenstein entropy 
\cite{ref45,ref46} can be recovered in this case. It is essential to note here that the entropy  in equation (1), is close to Tsallis' 
non-extensive entropy \cite{ref49b,ref49c}. In the case where the deformation effects are quantified with $\triangle$,  Barrow holographic 
dark energy will leave the regular one, leading to numerous cosmological variations. Recently, Barrow et al. \cite{ref49a} have used Big Bang 
Nucleosynthesis (BBN) data in order to impose constraints on the exponent of Barrow entropy. They have shown that the Barrow exponent 
should be inside the bound $\triangle \lesssim 1.4\times 10^{-4}$ in order not to spoil the BBN epoch. \\

Therefore, the BHDE is surely a more general structure than the standard HDE scenario. Here we concentrate on the general case of 
$(\triangle>0)$. If we assumed that the Hubble horizon $H^{-1}$ as the IR cutoff $(L)$, we can write the energy density of BHDE as
 
\begin{equation}
\label{3}
\rho_{_{D}}= C H^{2-\triangle}
\end{equation}

Let us consider a spatially flat, homogeneous and isotropic, FLRW universe  the standard metric is given by

\begin{equation}
\label{4}
ds^{2}=-dt^{2}+a^{2}(t)(dr^{2}+r^{2}d\Omega^{2})
\end{equation}
In a flat FLRW Universe, the field equations for BHDE are given as :

\begin{equation}
\label{5}
H^{2}=\frac{1}{3}8\pi G (\rho_{_{D}}+\rho_{m})\\
\end{equation}

where $ \rho_{_{D}}$  is the energy density of BHDE and $\rho_{m}$ is the energy

density of  matter respectively. The energy density parameter of BHDE and matter can be given as
$\Omega_{m}= \frac{8\pi\rho_{m}G}{3 H^{2}}$ and $\Omega_{_{D}}= \frac{8\pi\rho_{_{D}}G}{3 H^{2}}$.\\

We know that the relation
\begin{equation}
\label{6}
\Omega_{_{BD}}+\Omega_{m}=1
\end{equation}
The conservation law  BHDE and matter are defined as :
\begin{equation}
\label{7}
\dot\rho_{m}+3H\rho_{m}=0
\end{equation}
\begin{equation}
\label{8}
\dot\rho_{_{D}}+3H(p_{_{D}}+\rho_{_{D}})=0
\end{equation}
 
 From Eq. (\ref{3}), we get
 
 \begin{equation}
 \label{9}
 \dot\rho_{_{D}}= \frac{3C}{2} (2-\triangle)H^{2-\triangle} \left(\frac{\Delta \Omega _D}{(\Delta -2) \Omega _D+2}-1\right)
 \end{equation}

 Now,  Eqs. (\ref{5}), (\ref{7}) and (\ref{8}) and combining the outcome with the Eq. (\ref{6}), we obtained
 \begin{equation}
 \label{10}
 \frac{\dot{H}}{H^2}=\frac{3}{2} \left(\frac{\Delta \Omega _D}{(\Delta -2) \Omega _D+2}-1\right)
 \end{equation}
 The deceleration parameter $q$ is written as  
\begin{equation}
\label{11}
q=-1-\frac{\dot H}{H^{2}}
\end{equation}
By using Eq. (\ref{10}), the deceleration parameter $q$ is also written as 

 \begin{equation}
 \label{12}
q=\frac{1-(\Delta +1) \Omega _D}{(\Delta -2) \Omega _D+2}.
 \end{equation}

 By utilizing the Eq. (\ref{8}) with Eqs. (\ref{9}) and (\ref{10}), we get the expression for the EoS parameter derived as:
\begin{equation}
\label{13}
\omega_{D}=-\frac{\Delta }{(\Delta -2) \left(1-\frac{(z+1)^3 \Omega _{\text{m0}}}{-\Omega _{\text{m0}}+(z+1)^3 \Omega _{\text{m0}}+1}\right)+2},
\end{equation}
where dash is the derivative, here we differentiate the EoS  parameter $\omega_{_{D}}$ with respect to $lna$ then we get $\omega_{_{D}}^{'}$. 
By using the Eqs. (\ref{9}) and (\ref{10}), we find

\begin{equation}
 \label{14}
  \omega^{'}_D= -\frac{3 \Delta ^3 \left(\Omega _D-1\right) \Omega _D}{\left((\Delta -2) \Omega _D+2\right){}^3}.
 \end{equation}
Similarly by using the Eqs. (\ref{9}) and (\ref{10}), we obtained $\Omega_{_{D}}$ as:
\
 \begin{equation}
 \label{15}
 \Omega^{'}_D=-\frac{3 \Delta  \Omega _D \left(\Omega _D-1\right)}{(\Delta -2) \Omega _D+2}
 \end{equation}

%%%%%%%%%%%%%%%%%%%%%%%%%%%%   Figure 1  %%%%%%%%%%%%%%%%%%%%%%%%%%%%%%%%%%%%%%%%%%%%%%%%%%%%%%%%%%%%%%%%%%%%%

\begin{figure}[H]
	\centering
	\includegraphics[width=10cm,height=6cm,angle=0]{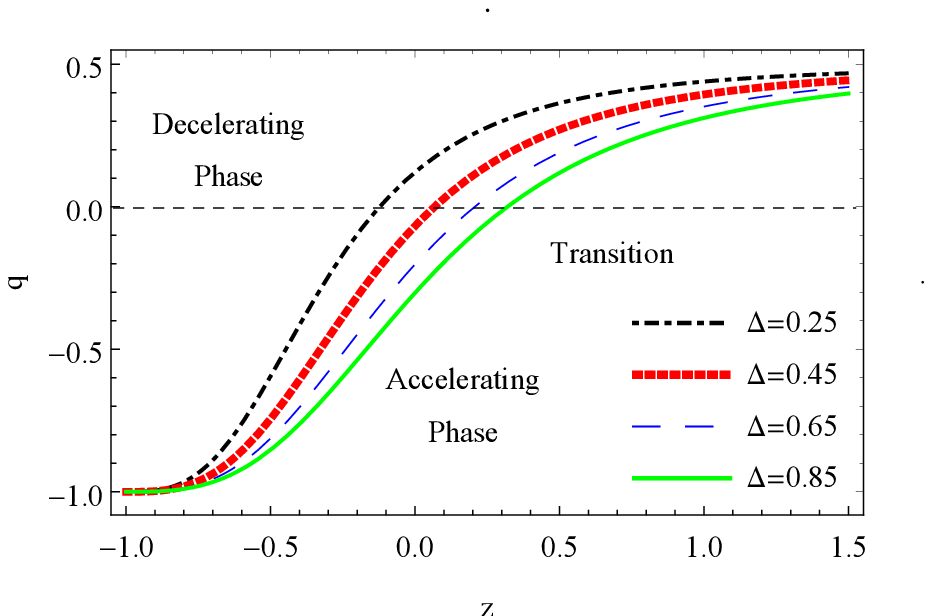}
	\caption{ Plot of deceleration parameter $(q)$ with redshift $z$}
\end{figure}
%%%%%%%%%%%%%%%%%%%%%%%%%%%%%%%%%%%%%%%%%%%%%%%%%%%%%%%%%%%%%%%%%%%%%%%%%%%%%%%%%%%%%%%%%%%%%%%%%%%%%%%%%%%%%%%
The evolution of $q$ has been plotted
in Fig. $1$. As we observed from Fig. $1$, the 
BHDE model can explain the universe's history very
well, with the sequence of an early matter-dominated era. Here we plot the  $q$ versus $z$ for a various choice of Barrow exponent $\triangle$. 
Which explains that the model is stable in the era of matter dominance.  Moreover, we analyzed that in the high redshift phase, we have $q\to-1$, 
while at $z\to-1$. It is worth mentioning that, cosmos may cross the phantom line ($q<-1$) for $z<-1$ depending on the value of $\triangle$.
The decelerating parameter approaches positive to negative values when the universe is overcome by dark energy. However, our findings based 
on the different values of the $\triangle$. If we take $\triangle = 0.25, 0.45, 0.65, 0.85$ the decelerating parameter $q$ is deceleration  
to accelerating for the present time. Additionally, the transition redshift $ z_{t} = 0$ occurs within the interval
$- 0.25 < z_{t} < 0.25$, which are in good compatibility with different
 recent studies (see Refs. \cite{ref50}-\cite{ref56} for more details
 about the models and cosmological datasets used). It has also been observed that the parameter $z_{t}$ depends on
 the values of $\triangle$ in such a way that, as  $\triangle$  increases, the
 parameter $z_{t}$ also increases. According to the Planck measurement of $\Omega_{D}$, the value of $r$ is $0.445\pm 0.010$. 
  
%%%%%%%%%%%%%%%%%%%%%%%%%%%%%%%%%%      Figure 2   %%%%%%%%%%%%%%%%%%%%%%%%%%%%%%
\begin{figure}[H]
\centering
\includegraphics[width=9cm,height=9cm,angle=0]{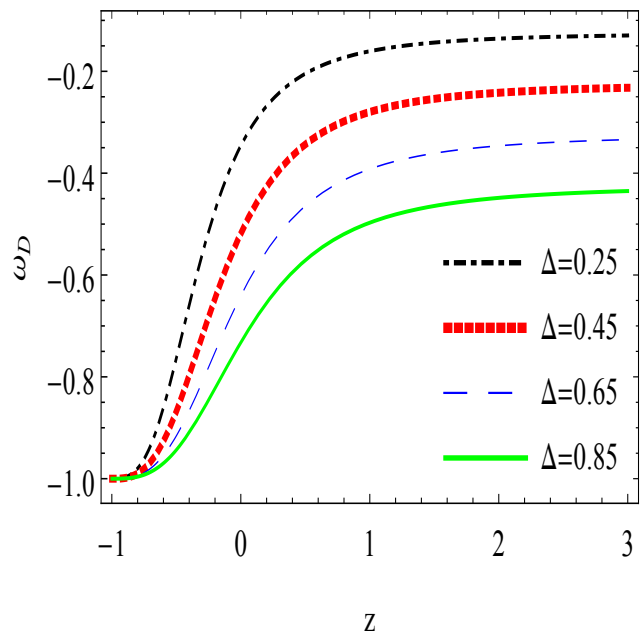}
\caption{ Plot of EoS parameter ($\omega_{_D}$) with redshift $z$}
\end{figure}

%%%%%%%%%%%%%%%%%%%%%%%%%%%%%%%%%%%%%%%%%%%%%%%%%%%%%%%%%%%%%%%%%%%%%%%%%%%%%%%%%%%%%%%%%%%%%%%%%%%%%

Next, we have shown the evolution of the EoS parameter $\omega_{_{D}}$ in Fig. $2$ by considering different values of $\triangle$ with 
respect to redshift $z$. The expression for equation of state parameter $\omega_{_{D}}$ represents in Eq. (\ref{13}). 
One of the main efforts in observational cosmology is the measurement of EoS for dark energy (DE).
Interestingly, we observed that for different values of $\triangle$, the EoS parameter $\omega_{_{D}}$ lies in the quintessence regime 
$(\omega_{_{D}}>-1)$ at the present epoch, however it enters in the phantom regime $(\omega_{_{D}}<-1)$  in the far future (i.e.,$z\to-1$). 
On the other hand, we also observed from the figure that the EoS parameter was very close to zero at high redshift and attains some $-ve$ 
value in between the region $-1$ to $-1/3$ at low redshift and further settles to a value very close to $-1$ in the far future. 
In this way, we see,  according to the value of $\triangle$, Barrow holographic dark energy can lie in the quintessence or in the 
the phantom regime, or exhibit the phantom-divide crossing during the cosmological evolution. \\

%%%%%%%%%%%%%%%%%%%%%%%%%%%%%%%%%%  firure 3  %%%%%%%%%%%%%%%%%%%%%%%%%5
\begin{figure}[H]
	\centering
	\includegraphics[width=9cm,height=9cm,angle=0]{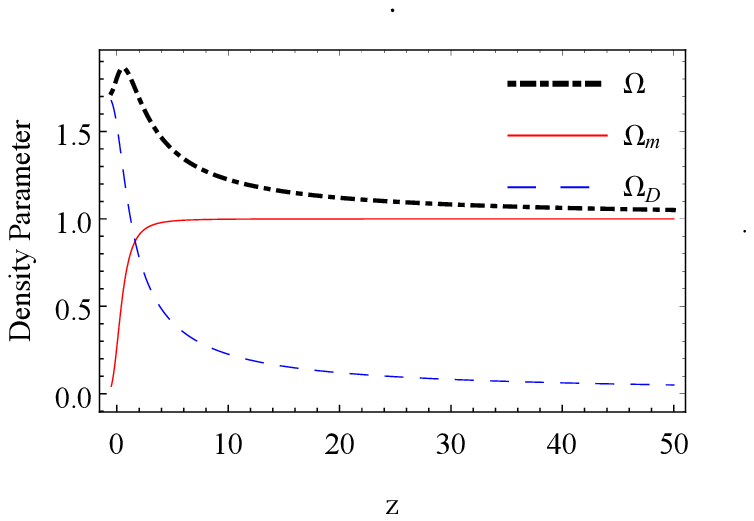}
	\caption{ Plot of density parameter ($\Omega_{_D}$) with redshift $z$}
\end{figure}
%%%%%%%%%%%%%%%%%%%%%%%%%%%%%%%%%%%%%%%%%%%%%%%%%%%%%%%%%%%%%%%%%%%%%%%%%%%%%%%%%%%%%%%%%

In this segment, we discuss cosmological development in the scenario of Barrow holographic dark energy. Figure $3$ shows the evolution
of the BHDE density parameter $\Omega_{_{D}}$ as a function of the redshift parameter $z$. From Fig. $2$, it is evident that $\Omega$ 
approaches unity as the universe evolves to high redshift, and Where $\Omega_{_{D}}$  is the density parameter for BHDE, 
and the $\Omega_{m}$ represents the density parameter of matter. By the assumption, \cite{ref57}, it has been seen that the current universe 
is near a spatially at geometry  ($\Omega \approx 1$). This really is a characteristic outcome from inflation in the early universe \cite{ref58}. 
Our figure depicts  that as $z \to 0$, $\Omega> 1$ or $\Omega< 1$ and when $z\to\infty$, $\Omega= 1$.

%%%%%%%%%%%%%%%%%%%%%%%%%%%%%%%%%%%%%%%%%%%%%%%%%%%%%%%%%%%%%%%% Section 3 %%%%%%%%%%%%%%%%%%%%%%%%%%%%%%%%%%%%%%%%%%%%%%%%%
\section{Statefinder}

In order to get a vigorous investigation to separate among DE models, many authors \cite{ref30,ref31} have presented a new mathematical 
diagnostic pair ${(r, s)}$, known as statefinder parameter, which is developed from the scale factor. These parameters ${(r, s)}$ is 
geometrical in the behavior and it is developed from the space-time metric directly. \\

The dynamics of the universe are comprehensively described by statefinder $(r, s)$. These are determined as

 \begin{equation}
\label{16}
r=\frac{\dddot a}{aH^{3}}
\end{equation}

\begin{equation}
\label{17}
s=\frac{(r-1)}{3(q-\frac{1}{2})}
\end{equation}

The relation between the statefinder parameters $r$ and $s$ in terms of energy density can be expressed as
\begin{equation}
\label{18}
r=\frac{(\Delta  (\Delta  (\Delta +12)-6)-8) \Omega _D^3+3 (\Delta  (4-7 \Delta )+8) \Omega _D^2+3 (\Delta -2) (3 \Delta +4) 
\Omega _D+8}{\left((\Delta -2) \Omega _D+2\right){}^3}
\end{equation}

\begin{equation}
\label{19}
s=-\frac{2 \left(\Omega _D-1\right) \left(2 (\Delta -1) \Omega _D-\Delta +2\right)}{\left((\Delta -2) \Omega _D+2\right){}^2}
\end{equation}

%%%%%%%%%%%%%%%%%%%%%%%%% Figure 4 (a) & (b)   %%%%%%%%%%%%%%%%%%%%%%%%%%%%%%%%%%%%%
\begin{figure}[H]
	\centering
(a)	\includegraphics[width=7cm,height=7cm,angle=0]{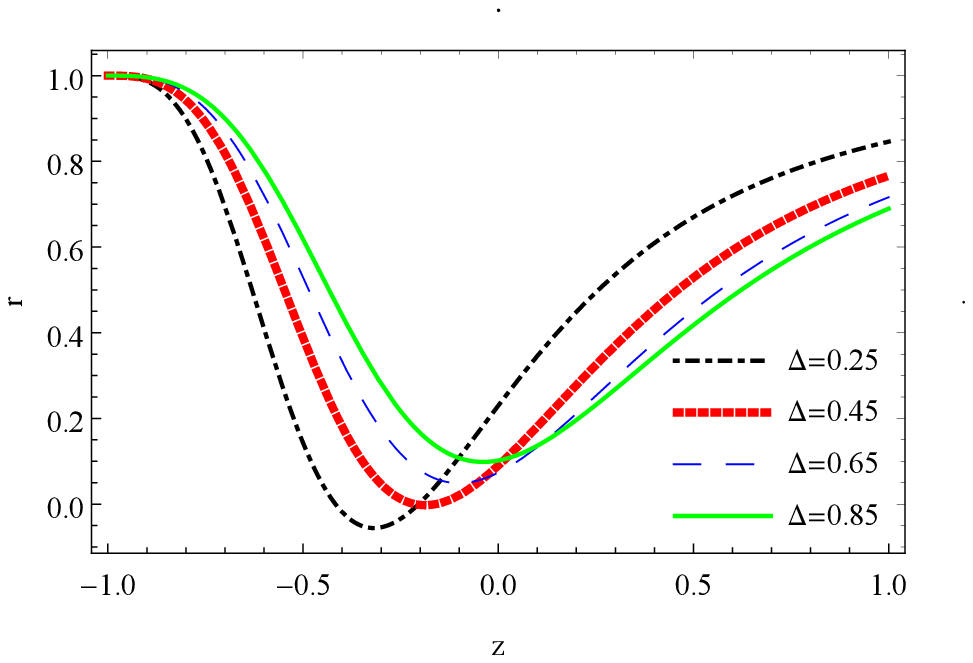}
(b)		\includegraphics[width=7cm,height=7cm,angle=0]{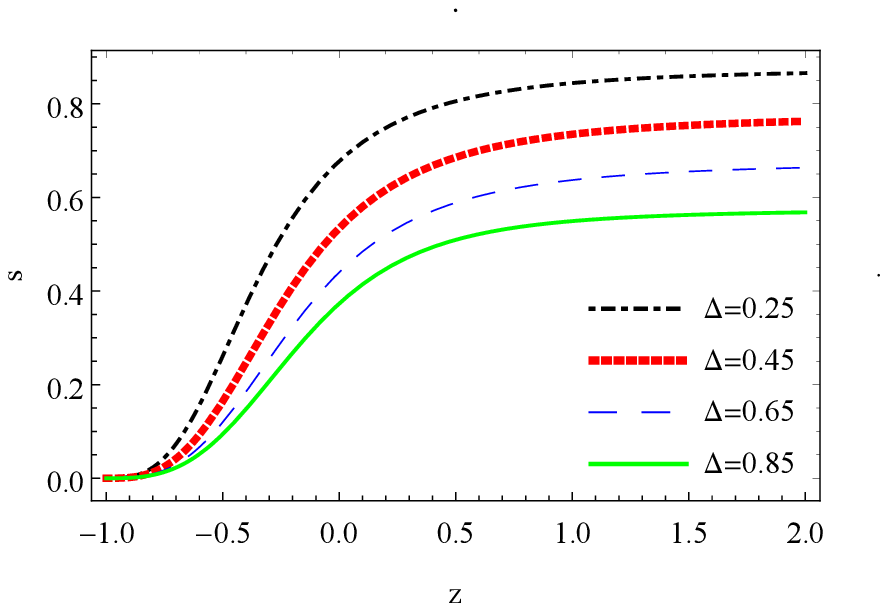}
	\caption{(a) Plot of $r$ with redshift $z$ (b)  Plot of $s$ with redshift $z$ }
\end{figure}
%%%%%%%%%%%%%%%%%%%%%%%%%%%%%%%%%%%%%%%%%%%%%%%%%%%%%%%%%%%%%%%%%%%%%%%%%%%%%%%%%%%%%%%%%%%%%%%%%%%%%%
The evolution of $r$ and $s$ with redshift $z$ for FLRW universe has been analyzed in Figs. $4a$ and $4b$ \cite{ref59}. The primary 
parameter $r$ of Oscillating dark energy (ODE), at high redshift, approaches standard $\Lambda CDM$ behavior while at low redshift it 
goes deviates significantly from the standard behavior and the second parameter $s$ shows opposite in behavior \cite{ref60}.
Figures $4a$ and $4b$ portray evaluation of $r$ and $s$ for different values of Barrow parameter $\triangle$ and approaches to 
the $\Lambda CDM$, by taking the value (for $\Omega_{m_{0}}$ = $0.27$ and $H_{0}$= $69.5$) are in good arrangement with recent observations. 
As expected, for $\triangle= 0$ the above modified Friedmann equations reduce to $\Lambda CDM$ scenario.
 The study of the statefinder provides a very useful method to split the conceivable depravity of different cosmological models by 
 determining the parameters $r$, and $s$ for the higher order of the scale factor.

%%%%%%%%%%%%%%%%%%%%%%%%%%%%%% figure 5 (a) & (b) %%%%%%%%%%%%%%%%%%%%%%%%%%%%%%%%%%%%%%%%%%%%%%%%%%%%%%%%%%%%
\begin{figure}[H]
	\centering
	(a)\includegraphics[width=7cm,height=7cm,angle=0]{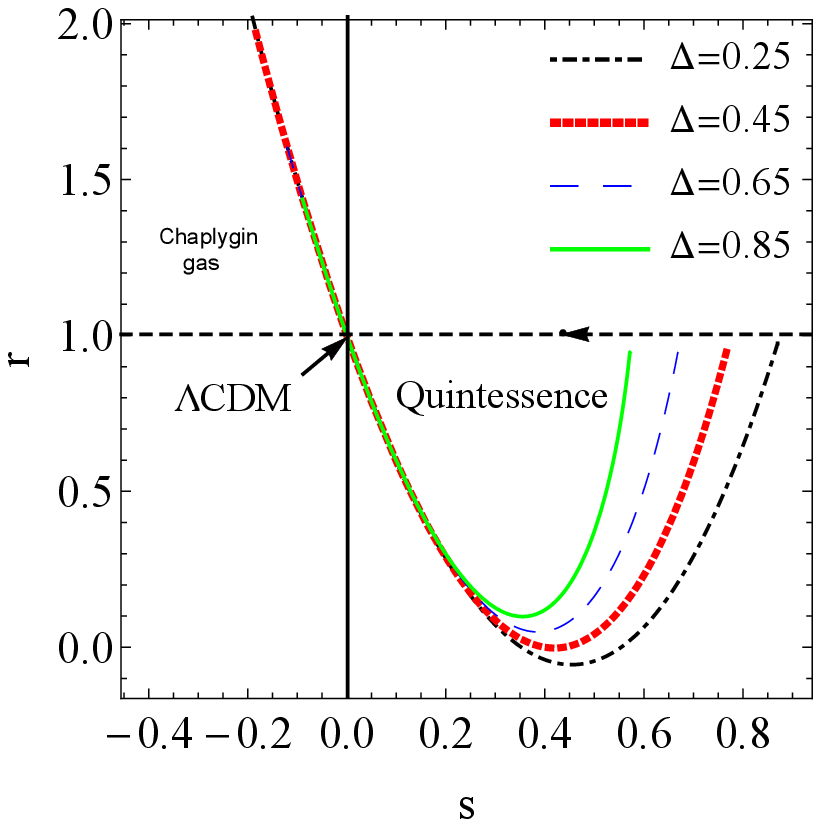}
	(b)\includegraphics[width=7cm,height=7cm,angle=0]{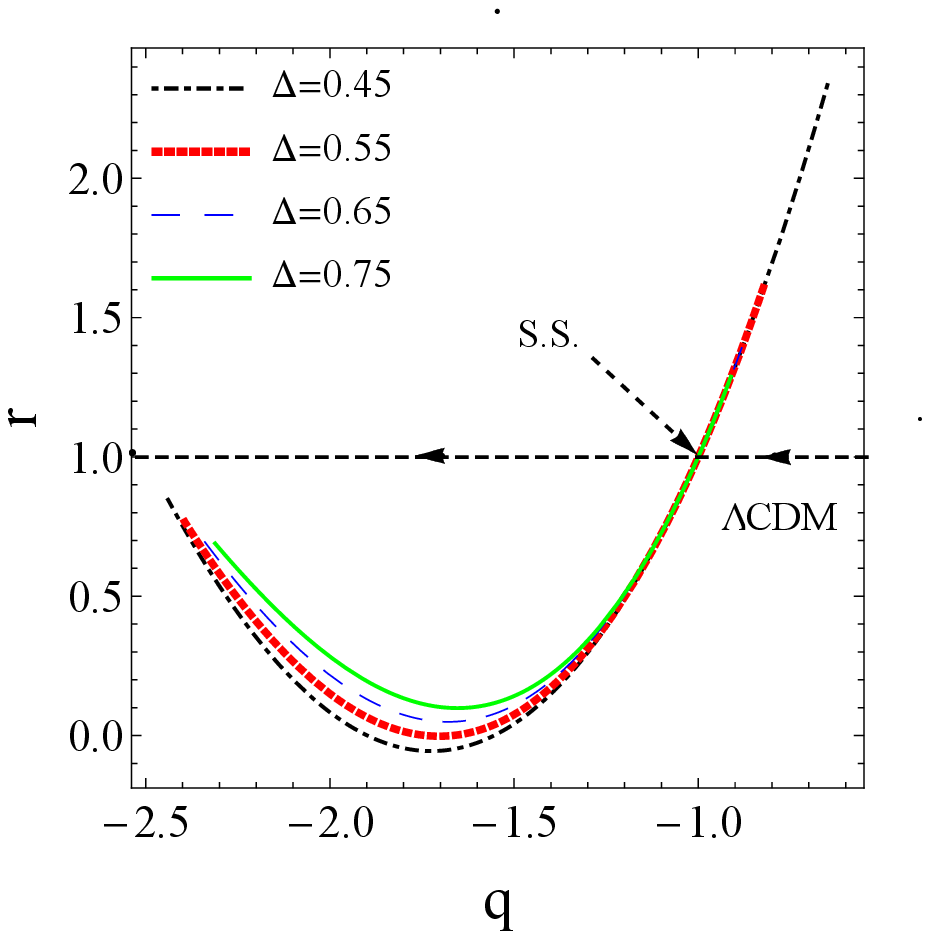}
	\caption{(a) Plot of $(r - s)$ with $z$ (b) Plot of $(r - q)$ with  $z$ }
\end{figure}
%%%%%%%%%%%%%%%%%%%%%%%%%%%%%%%%%%%%%%%%%%%%%%%%%%%%%%%%%%%%%%%%%%%%%%%%%%%%%%%%%%%%%%%%%%%%%%%%%%%%%%%%%%

In Fig. $5a$, we plot  $r-s$ trajectories which divided into two regions. The region $r>1$, $s<0$ in the $r-s$ plane, shows a 
behaviour similar to a Chaplygin gas (CG) model \cite{ref61} whereas the region $r<1$,  $s > 0$ shows a behaviour similar to the 
quintessence model (Q- model) \cite{ref30,ref31}. However, the model shows a behavior of CG at early time for $\triangle=0.25,0.45,0.65,0.85$ 
and approaches $\Lambda CDM$  at late times. The trajectories in both region coincide for all different values of $\triangle$. The statefinder 
${r,s}$ of BHDE model approaches to the $\Lambda CDM$. In addition, we also plot the evolution trajectory in the $r-q$ plane in figure $5b$. 
The $(r-q)$ trajectories are divided into two regions through the point $(r,q)=(1,-1)$. The region $r>1$, $q<-1$ in the $r-q$ plane shows a 
behaviour similar to the phantom model, while the region $r<1$ $q>-1$ shows a behaviour similar to quintessence (Q-model). In this figure, 
the arrow represents the fixed points at ${r,q}={1,-1}$ of the steady-state (SS) model. As exhibited in \cite{ref36}-\cite{ref48} the 
statefinder can effectively separate between a wide assortment of DE models including the quintessence, phantom, quintom,  cosmological 
constant, braneworld models, Chaplygin gas, and interacting DE models. 
  
 %%%%%%%%%%%%%%%%%%%%%%%%%%%%%%%%%%%%%%%%%%%%%%%%%%%%%%%%%%% Section 4 %%%%%%%%%%%%%%%%%%%%%%%%%%%%%%%%%%%%%%%%%%%%%%%%%%%%%%%%%%%%%%%%% 
  
  \section{$O_{m}(z)$ Diagnostics}
  The Om diagnostic analysis is also a  very useful geometrical diagnosis that can be used for such analysis. In the investigation of 
  the statefinder parameter  $(r, s)$ the higher order derivative of $a(t)$ are utilized. The Ist order derivative are used in $O_{m}$ 
  diagnostic analysis because it contains only the Hubble parameter.\\
  
  The $O_{m}$ diagnostic can be considered as an easier diagnostic \cite{ref62}. It might be noticed that the $O_{m}$ diagnostic has 
  been additionally  applied to \cite{ref63}-\cite{ref65}. This arrangement of parameters can be written as:

 \begin{equation}
\label{20}
Om(z)= \frac{\frac{H^{2}(z)}{H_{0}}-1}{(1+z)^{3}-1}
\end{equation}
%%%%%%%%%%%%%%%%%%%%%%%%%%%%%%%%%%%%%%%%%%%%%%%%%%%%%%%%%%%%%%%%%% Figure 6 %%%%%%%%%%%%%%%%%%%%%%%%%%%%%%%%%%%%%
\begin{figure}[H]
	\centering
	\includegraphics[width=9cm,height=9cm,angle=0]{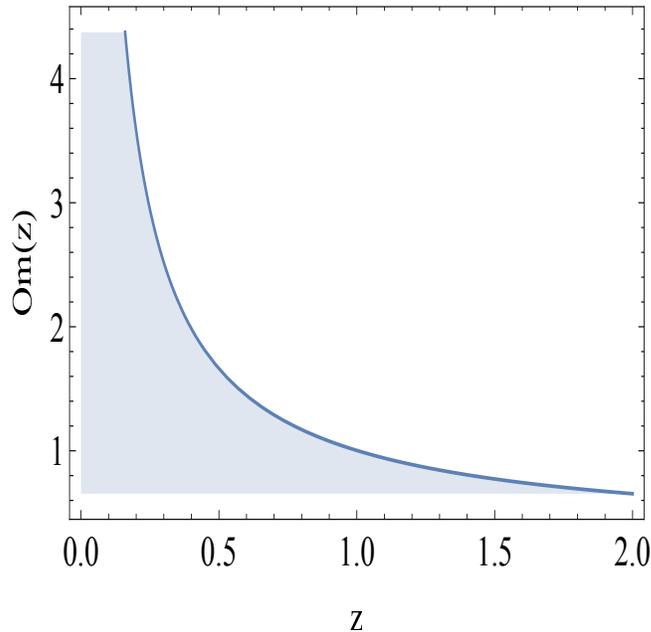}
	\caption{ Plot of $O_{m}(z)$ with redshift $z$}
\end{figure}
%%%%%%%%%%%%%%%%%%%%%%%%%%%%%%%%%%%%%%%%%%%%%%%%%%%%%%%%%%%%%%%%%%%%%%%%%%%%%%%%%%%%%%%%%%%%%%%%%%%%%%%%%%%%%%%%%%%%%%%% 
 In figure $6$, we plot the $O_{m}(z)$ evolution with redshift.
The positive curve  of the  $O_{m}(z)$ trajectories shows the  phantom  behaviour $(\omega <-1)$ whereas the negative curve 
implies that DE behaves like quintessence $(\omega >-1)$. We noticed in our figure, when the redshift z is expanding inside the stretch 
$ 0 < z < 2.0$, the $O_{m}(z)$ is diminishing monotonically and the curve lies in phantom region. The new diagnostic of dark energy $O_{m}$ 
is acquainted with separate $\Lambda CDM$ from other DE models. Where $H_{0}$ is the current estimation of the Hubble parameter. Here we 
demonstrated that the slope of $O_{m}(z)$ can recognize dynamical DE from the cosmological consistent in a robust way.
  
 %%%%%%%%%%%%%%%%%%%%%%%%%%%%%%%%%%%%%%%%%%%%%%%%%%%%%%%%%%%%%%%%% Section 5 %%%%%%%%%%%%%%%%%%%%%%%%%%%%%%%%%%%%%%%%%%%%%%%%%%%%
  \section{Conclusion}
  
In this model, we have discussed the BHDE, by considering the typical holographic principle at a cosmological system, by utilizing 
the Barrow entropy, rather than the standard Bekenstein-Hawking. Here we have also discussed the evolution of a spatially
flat FLRW universe composed of pressure less dark matter and Barrow holographic dark energy. By considering the Hubble horizon 
as the infrared cut-off, we have found the exact solution and the calculated cosmological parameters like the behavior of the density
parameter, the EoS parameter, the deceleration parameter, statefinder, and $O_{m}$ diagnostic parameters, etc. We also plotted the 
trajectories in $(r-s), (r-q)$, and $O_{m}$  to discriminate the various DE model from the existing BHDE models during 
the cosmic evolution.\\

The main highlights of the models are as per the following:
 
 \begin{itemize}
 	\item  It has been found that the BHDE model exhibits a
 	smooth transition from early deceleration era $(q> 0)$ to
 	the present acceleration $(q< 0)$  era of the universe in Fig. $1$.  Also,
 	the value of this transition redshift is in good accordance
 	with the current cosmological observations and obtained for the different values of the $\triangle$.
 	
 	\item 
 	 It has  been  observed in Fig. $2$ that the new Barrow exponent $\triangle$ essentially influences the dark energy equation 
 	 of state and as per its worth it lies in the quintessence regime $(\omega_{D}>-1)$, at the current era, however it enters in 
 	 the phantom regime $(\omega_{D}<-1)$  in the far future (i.e.,$z\to-1$)
 	by using different values of $\triangle$. 
 	
 	\item 
 	 The energy density parameter is also discussed and shown in Fig. $3$. We found that in the cosmic evaluation of BHDE $\Omega$, approaches unity. 
 	 which is a good agreement with recent observations.
 	 
 	\item 
 	We have also discussed the statefinder $(r, s)$  in terms of the dimensionless density parameters and Barrow exponent $\triangle$. 
 	We have plotted  $r$ verses $z$ in Fig. $4a$. The  $r(z)$ parameter of oscillating dark energy (ODE) depicts in  high red shift region 
 	and  approaches to standard $\Lambda CDM$. Similarly we have obtained $s(z)$  in Fig. $4b$, where $s(z)$ parameter shows opposite behaviour 
 	to the primary parameter $r$.
 	
 	\item 
 	 The excellent diagnostics of DE is shown in Figs. $5a$ and $5b$ which are ${(r, s)}$ and ${(r, q)}$.  Here we take the value  
 	 $\Omega_{m_{0}}=0.27$, $H_{0}=69.5$ and  using the different values of $(\triangle= 0.25, 0.45, 0.65,0.85)$, then the 
 	 averaged-over-redshift statefinder pair ${(r,  s)}$ and ${(r, q)}$ obtained the Chaplygin gas (CG) model, steady state (SS) model, 
 	 quintessences (Q-model) etc. Now we observe that the statefinders play a very important role in the FLRW universe with BHDE.
 	
 	\item 
 	The $O_{m}$-diagnostic technique is used to check the stability of the model and various periods of the Universe. We plot
 	the trajectory in Om(z)plane to separate the conduct of the DE models in Fig. $6$. The positive inclination  of the curve shows 
 	the phantom-like behavior of the model
 	
 \end{itemize}  
  In summary, in the manuscript, the physical behavior of the cosmological parameters are studied through their graphical representation. 
 This BHDE model is in a good agreement with cosmological informations, and it can fill in as a decent possibility for the graphical representation.

  %%%%%%%%%%%%%%%%%%%%%%%%%%%%%%%%%%%%%%%%%%%%%%%%%%%%%%%%%%%%%%%%%%%%%%%%%%%%%%%%%%%%%%%%%%%%%%%%%%%%%%%%%%%%%%%
\section{Acknowledgments}
The author (AP) thanks the IUCAA, Pune, India for providing the facility under visiting associateship. The authors are also thankful to 
the anonymous referee for his/her constructive comments which helped to improve the quality of paper in present form.
%%%%%%%%%%%%%%%%%%%%%%%%%%%%%%%%%%%%%%%%%%%%%%%%%%%%%%%%%%%%%%%%%%%%%%%%%%%%%%%%%%%%%%%%%%%%%%%%%%%%%%%%%%%%%%%%

\end{document}